\documentclass[letterpaper,12pt]{article}

\usepackage{endnotes}
\usepackage{amssymb}
\usepackage{amsfonts}
\usepackage{amsmath}
\usepackage{amsthm}
\usepackage{graphicx,color,xcolor}
\definecolor{db}{rgb}{0.2, 0.46, 0.63}

\usepackage{comment}
\usepackage{array}

\usepackage{url}

\usepackage{setspace}

\usepackage{natbib}
\setcitestyle{numbers}
\setcitestyle{square}

\def\R{\mathbb{R}}
\def\endproof{\hfill\diamondsuit}

\def\sF{{\mathcal F}}

\def\sA{{\mathcal A}}
\def\sL{{\mathcal L}}

\def\sC{{\mathcal C}}

\def\E{\mathbb{E}}

\def\sF{\mathcal{F}}
\def\P{\mathbb{P}}

\def\N{\mathbb N}

\numberwithin{equation}{section}
\theoremstyle{plain}                
\newtheorem{theorem}{Theorem}[section]
\newtheorem{lemma}[theorem]{Lemma}

\theoremstyle{definition}           
\newtheorem{definition}[theorem]{Definition}

\theoremstyle{remark}               

\usepackage[margin=1.25in]{geometry}

\usepackage{ulem} 

\begin{document}

\begin{center}
\large{\bf Long-run survival in limited stock market participation models with power utilities}\footnote{ Kasper Larsen is corresponding author and has contact information: Email: \url{KL756@math.rutgers.edu} and mailing address: Department of Mathematics, Rutgers University, Hill Center 330 - Busch Campus, 110 Frelinghuysen Road, Piscataway, NJ 08854-8019, USA. Phone: +1.848.445.2390. We thank Paolo Guasoni for discussions and we thank our three anonymous reviewers for their time and useful comments.}

\ \\

{\large \bf Heeyoung Kwon}\\
Rutgers University

\ \\

{\large \bf Kasper Larsen}\\
Rutgers University

\ \\

\end{center}
\begin{center}
\ \\

{\normalsize \today }
\end{center}
\vspace{.5cm}

\begin{verse} {\sc Abstract}:  We extend the limited participation model in Basak and Cuoco (1998) to allow for  traders with different time-preference coefficients but identical constant relative risk-aversion coefficients. Our main result gives parameter restrictions which ensure the existence of a Radner equilibrium. As an application, we give further parameter restrictions which ensure all traders survive in the long run. 
\end{verse}

\vspace{0.25cm}

\newpage

\noindent Declaration of interest: Heeyoung Kwon has no conflicts of interest.  Kasper Larsen has no conflicts of interest.  \ \\

\noindent Declaration of generative AI in scientific writing: The original submission was done without any AI usage. {\color{black} The authors used ChatGPT to obtain feedback on exposition and to
check mathematical arguments during the revision process. The authors
independently verified the resulting content and assume responsibility
for all content.}\ \\

\noindent Data availability statement: We do not analyze nor generate any datasets.

\newpage

\section{Introduction}

Basak and Cuoco (1998) construct a continuous-time Radner equilibrium model with two traders. One trader can access both the stock and money markets whereas the second trader cannot hold stocks. When both traders have log utilities and identical time preferences, Basak and Cuoco (1998) prove existence of a  Radner equilibrium. We consider a model extension where both traders have identical power-utility functions but have different time-preference coefficients. Our main result gives parameter restrictions which ensure the existence of a Radner equilibrium. Our existence proof is based on showing that a non-linear, singular, and path-dependent first-order ODE has a global $\sC^1$ solution. As an application, we show that different time-preference parameters can produce long-run survival of both traders.

There exist several variations of Basak and Cuoco (1998). For example, Hugonnier (2012)  considers more general participation constraints and proves existence of bubbles in equilibrium (i.e., the stock price differs from its discounted future dividends).  Prieto (2013) extends Hugonnier (2012) further and proves the existence of an equilibrium when the unrestricted trader has a power-utility function. Both Hugonnier (2012) and  Prieto (2013) assume that the restricted trader has a log-utility function. Because the restricted trader faces an incomplete financial market, her optimal investment and consumption problem is difficult to analyze, however, the log-utility assumption makes it explicitly solvable. Finally, we mention Weston (2024) who proves equilibrium existence  for traders with heterogeneous exponential utilities. Because exponential utilities have domain $\R$, Weston (2024) can allow consumption rates to become negative and this model relaxation makes the exponential optimization problems non-singular. 

More recently,  Guasoni, Larsen, and Leoni (2025) prove equilibrium existence when both traders have identical power utilities and identical time-preference coefficients. Their equilibrium existence proof  hinges on proving global existence of a $\sC^1$ solution to a  non-linear, singular, and path-dependent first-order ODE with quadratic growth terms. We extend their setting to allow for different time-preference coefficients (but keeping identical power utilities). {\color{black} More specifically, we assume the restricted trader is more patient and prove existence under an additional bound on the difference of the traders' time-preference parameters}. Our relaxation produces a new cubic term in the resulting ODE and our main mathematical contribution is to prove that a unique global $\sC^1$ solution exists of the resulting ODE. 

We conclude by applying our equilibrium to studying long-term survival of traders. In asset pricing theory, a model's  stability properties are often used to judge its quality.\footnote{For example, to resolve asset pricing puzzles, such as the interest rate puzzle from Weil (1989)  and the equity premium puzzle from Mehra and Prescott (1985), many popular models exhibit stationarity properties because such properties are useful for model calibration.}  In continuous-time settings similar to ours, Kogan, Ross, Wang, and Westerfield (2006) show that traders with incorrect beliefs have zero long-run consumption-share limits and Yan (2008)  shows that a trader’s long-run consumption share limit is determined by her survival index.  More recent references on survival analysis include Bhamra and Uppal (2014),  Borovi\v{c}ka (2020), and  Huang and  Liu (2025).  Unfortunately, the equilibrium models in both Basak and Cuoco (1998) and  Guasoni, Larsen, and Leoni (2025) cannot produce survival of the restricted trader because the long-run limit of her  consumption-share process is zero. In contrast, we show that our model with  heterogeneous time-preference parameters allows for both traders to survive in the long run. For comparison, we show that the log-power model in Prieto (2013) can also produce surviving  traders.


\section{Radner equilibrium }\label{sec:application}

The following modeling setting is from Basak and Cuoco (1998). To study the traders' survival properties, we use an infinite time horizon. However, this model variation of  Basak and Cuoco (1998) is not new and is already discussed in Remark 3 in Hugonnier (2012). Our probability space is denoted by $(\Omega,\mathbb{F},\P)$ on which $(B_t)_{t\ge0}$ is a Brownian motion. The filtration is $\sF_t^0 := \sigma(B_s)_{s\in [0,t]}$ and we assume  $\mathbb{F} = \vee_{t\ge0} \sF_t^0$. As usual, the augmented filtration is defined as $\sF_t:=\sF_t^0\vee\mathbb{F}$'s $\P$-nullsets for $t\ge0$.

\subsection{Individual optimization}

We consider a pure-exchange economy where the consumption good serves as the model's num\'eraire. The single stock pays dividends at rate $D =(D_t)_{t\ge0}$ where
\begin{align}\label{dD}
dD_t := D_t \big(\mu_D dt + \sigma_D dB_t\big),\quad D_0>0.
\end{align}
In \eqref{dD}, the constants $D_0>0$, $\mu_D\in\R$, and $\sigma_D>0$ are model input. The stock has price processes $S=(S_t)_{t\ge0}$ and money market account has price process $S^{(0)}=(S^{(0)})_{t\ge0}$. These two price processes are conjectured to have It\^o dynamics 
\begin{align}
dS^{(0)}_t &= r_t S^{(0)}_t dt, \quad S_0^{(0)}:=1,\label{dS0}\\
dS_t &= (S_t r_t -D_t)dt +S_t \sigma_{S,t} (\kappa_{t}dt+dB_t),\quad S_0\in (0,\infty).\label{dS}
\end{align}
In \eqref{dS0} and \eqref{dS}, the quantities $S_0\in (0,\infty)$, $r \in \sL^1_\text{loc}$, and $(\kappa,\sigma_{S})\in \sL^2_\text{loc}$ are determined in equilibrium. For notational simplicity, we normalize the shares of stock to one and the shares of the money market to zero.

Trader 1 can trade both the stock and the money market and her wealth process has dynamics 
\begin{align}\label{dW1}
\begin{split}
dX_{1,t} & := r_t X_{1,t} dt +  \theta_{1,t}S_{t}  \sigma_{S,t}(\kappa_{t}dt+ dB_t) - c_{1,t}dt,\\ 
X_{1,0}&:= \theta^{(0)}_{1,0-} +S_0\in\R.
\end{split}
\end{align}
In \eqref{dW1}, trader 1's controls are the consumption rate $c_1$ and the number of shares held $\theta_1$. As in  Basak and Cuoco (1998), trader 1's endowed shares of the money market account  $\theta^{(0)}_{1,0-}$ is assumed to be negative, i.e.,  $\theta^{(0)}_{1,0-} \in (-\infty,0)$.

Trader 2 cannot trade the stock and so her wealth process has dynamics 
\begin{align}\label{dW2}
\begin{split}
dX_{2,t} & := r_tX_{2,t} dt  - c_{2,t}dt,\\ 
X_{2,0}&:= \theta^{(0)}_{2,0-}\in(0,\infty).
\end{split}
\end{align}
In \eqref{dW2}, trader 2's control is the consumption-rate  process  $c_2$ and $\theta^{(0)}_{2,0-}=- \theta^{(0)}_{1,0-} >0$ denotes her   number of endowed shares of the money market account. 

Compared to  Guasoni, Larsen, and Leoni (2025), our model allows for different time-preference coefficients $\beta_1>0$ and $\beta_2>0$. However, the two traders have a common constant relative risk-aversion coefficient $\gamma \in (0,1)$. We assume that trader 1 has objective 
\begin{align}\label{primalval}
&\sup_{\theta_1,c_1\in \sA_1}\tfrac{1}{1-\gamma}\E\left[\int_0^\infty e^{-\beta_1t}c^{1-\gamma}_{1,t}dt\right]
\end{align}
whereas trader 2 has objective 
\begin{align}\label{primalval1}
&\sup_{ c_2\in \sA_2}\tfrac{1}{1-\gamma}\E\left[\int_0^\infty e^{-\beta_2t}c^{1-\gamma}_{2,t}dt\right].
\end{align}
 As we shall see in Section \ref{survival} below, this extension allows both traders to survive in the long run. In \eqref{primalval} and \eqref{primalval1}, the admissible sets $\sA_1$ and $\sA_2$ in \eqref{primalval} and \eqref{primalval1} are defined as:

 \begin{definition}[Admissibility] \label{def:ad} Progressively measurable processes $(\theta_1,c_1)$ are  admissible  iff $c_{1,t}\ge0$ for all $t\ge0$ and the solution of \eqref{dW1} exists and  satisfies $X_{1,t}\ge0$ for all $t\ge0$. In this case, we write $(\theta_1,c_1) \in \sA_1$. Similarly, a progressively measurable process $c_2$ is admissible iff $c_{2,t}\ge0$ for all $t\ge0$ and the solution of \eqref{dW2} exists and satisfies $X_{2,t}\ge0$ for all $t\ge0$. In this case, we write  $c_2 \in \sA_2$.
$\endproof$
\end{definition}

\subsection{Equilibrium}
The following definition is standard and can be found in, e.g., Chapter 10 Duffie (2001). 
  
\begin{definition}[Radner equilibrium] \label{def:eq} A constant $S_0\in(0,\infty)$, progressively measurable processes  $r \in \sL^1_\text{loc}$ and $(\kappa,\sigma_{S})\in \sL^2_\text{loc}$, and controls $(\hat \theta_1, \hat c_1) \in \sA_1$ and $\hat c_2\in \sA_2$ constitute a Radner equilibrium iff:
\begin{itemize}
\item[(i)] The controls $(\hat{\theta}_1,\hat{c}_1)\in \sA_1$ maximize \eqref{primalval}.
\item[(ii)] The control $\hat{c}_2\in \sA_2$ maximizes \eqref{primalval1}.
\item[(iii)] The stock and consumption markets clear in the sense
\begin{align}
\hat{\theta}_{1,t} =1\quad \text{and }\quad \hat{c}_{1,t}+\hat{c}_{2,t}=D_t,\quad \text{for all }t\ge0.\label{clearingconds}
\end{align}
\end{itemize}
\vspace{-0.5cm}
$\endproof$
\end{definition}

\noindent{Walras' law ensures that clearing in both the stock and consumption markets implies that money market clears too. This additional clearing property stems from the self-financing wealth dynamics \eqref{dW1} and \eqref{dW2}.

{\color{black} Before giving the precise conditions under which we can prove existence, we heuristically discuss how to derive a candidate Radner equilibrium. We will create deterministic functions $\mu_Y,\sigma_Y:(0,1)\to \R$ and a corresponding  autonomous state-process $Y_t$ valued in $(0,1)$ with dynamics
\begin{align}\label{inhomeY}
dY_t = \mu_Y(Y_t)dt +\sigma_Y(Y_t) dB_t,\quad Y_0 \in (0,1).
\end{align} 
This process will be trader 1's consumption-share process in the sense that $\hat c_{1,t} :=D_tY_t$ maximizes \eqref{primalval}. For the equilibrium to be governed by a one-dimensional state process, it is crucial that the dividend-rate process in \eqref{dD} is a geometric Brownian motion.\footnote{\color{black}For example, had $D$ instead been a square-root process like $dD_t = a(b - D_t) dt + \sigma_D\sqrt{D_t} dB_t$, we would get a coupled two-dimensional state process. Our analysis cannot handle such an extension. } Based on the process $Y$, we express the equilibrium interest rate process as $r_t = r(Y_t)$ and the equilibrium market price of risk process as $\kappa_t = \kappa(Y_t)$ for suitable deterministic functions $r,\kappa :(0,1)\to \R$. Given these functions, the stock volatility $\sigma_S$ will be produced by the martingale representation theorem.\footnote{\color{black}Our analysis does not rely on $\sigma_S\neq0$. We leave it open to settle whether or not the unrestricted trader faces an endogenously complete market.}

To heuristically derive the functions $(\mu_Y, \sigma_Y, r,\kappa)$, we start by conjecturing that the unrestricted trader's maximizer $D_tY_t$ for \eqref{primalval} satisfies the first-order condition
\begin{align}
\begin{split}
e^{-\beta_1 t}(D_tY_t)^{-\gamma} &= \xi_{1,t},\quad d \xi_{1,t}= - \xi_{1,t}\big( r_t dt + \kappa_t dB_t\big),\quad \xi_{1,0}>0,\\
\end{split}
\end{align}
By applying It\^o's lemma and matching $dt$ and $dB$ coefficients, we produce expressions for $r$ and $\kappa$ in terms of $(\mu_Y,\sigma_Y)$. In turn, to get $(\mu_Y,\sigma_Y)$, we conjecture that the restricted trader's value function for \eqref{primalval1} has the form
$$
\frac{x^{1-\gamma}}{1-\gamma} u(y)^{-\gamma}, \quad u(y) := \frac{\xi}2 \exp\Big\{ \int_0^y \frac{h(q)-1}{1-q}dq\Big\},\quad y\in(0,1),
$$
for a constant $\xi>0$. This conjecture and the HJB equation give $\hat{X}_{2,t}u(Y_t)$ as the optimal consumption rate for trader 2 where $d\hat{X}_{2,t} := \hat X_{2,t}\big( r_t-u(Y_t)\big)dt$ as in  \eqref{dW2}.  By matching dynamics of $\hat{X}_{2,t}u(Y_t)$ with the dynamics of $D_t(1-Y_t)$, we produce two additional requirements that give  $(\mu_Y,\sigma_Y)$. All that remains is to determine $\xi$ and $h$, which is the content of the next lemma.
}

The next lemma gives the existence of a solution to a governing ODE, which we subsequently use to produce a Radner equilibrium. The lemma is proven in the next section.  {\color{black} We do not claim that our parameter restrictions are minimal. We  use the parameter restriction $\delta <0$ to ensure that the sign of the cubic term $h(y)^3$ in the below ODE \eqref{hODE} is positive.}

\begin{lemma}\label{MainLemma}   Let $\gamma \in (0,1)$, $\sigma^2_D>0$, $A \in (1+\delta - \frac{2\delta}{\gamma},\infty)$, and $\delta \in (-\gamma,0)$. 
\begin{enumerate}

\item There exists $\xi_0 \in(0,\infty)$ such that \eqref{hODE}-\eqref{hODEb} has a unique solution $h\in \sC^1([0,1])$, where 
 \begin{align}\label{hODE}
h'(y)  =a_0(y)+ \frac{a_1(y)}{1-y}h(y)+\frac{a_2(h,y)}{1-y}h(y)^2 + \frac{\delta y}{1-y}h(y)^2\Big ( 1- \frac{h(y)}\gamma\Big), 
\end{align}
for $ y\in(0,1)$,  boundary values
 \begin{align}\label{hODEb}
h(0) = \gamma,\quad h(1) =1,
\end{align}
and functions
\begin{align}\label{a0a1}
\begin{split}
a_0(y)&:= \frac{\gamma (1+\gamma )}{y},\quad y \in (0,1], \\
 a_1(y)& :=  \frac{(2 \gamma +1) y-(1+\gamma)}{y},\quad y \in (0,1],\\
a_2(h,y)& := \frac{\xi_0}{\sigma_D^2}\exp\Big\{\int_0^y \frac{h(q)-1}{1-q}dq\Big\}-A,\;\; y \in [0,1).
\end{split}
\end{align}
Furthermore, the solution satisfies  $\gamma \le h_{\xi_0}(y) \le 1$ for all $y \in [0,1]$ and  $h'(1) =\frac{(1-\gamma) (\gamma ^2+\gamma -\delta )}{\gamma  (A-\delta -1)+2 \delta}>0$.

\item For  the drift function $\mu_Y$ and volatility function $\sigma_Y$ defined as 
\begin{align}\label{mgmethod1bbq}
\begin{split}
\mu_Y(y)&:=\sigma_D^2 (1-y) \frac{y h(y) \big(2 \gamma ^2+\delta  y h(y)\big)-\gamma  (\gamma +1) (2 y-1)}{2 \gamma  y h(y)^2},\\
\sigma_Y(y)&:= \sigma_D\frac{1-y}{h(y)},
\end{split}
\end{align}
for $y \in (0,1)$, there exists a unique strong solution $Y$ of the SDE \eqref{inhomeY}. 
{\color{black} Furthermore, we have the boundary classifications:
\begin{enumerate}

\item The boundary point $y=0$ is inaccessible, entrance, and not attracting.

\item The boundary point $y=1$ is inaccessible and natural. For $\delta \in (-\gamma, -\gamma^2]$, $y=1$ is not attracting. For $\delta \in (-\gamma^2,0)$, $y=1$ is attracting.
\end{enumerate}
}

\end{enumerate}

\end{lemma}

The next theorem is our main contribution and it uses the function 
\begin{align}\label{g}
g(y):= \frac2{\xi_0} \exp\Big\{ -\int_0^y \frac{h(q)}{1-q}dq\Big\}(1-{\color{black}y})^{-\gamma},\quad y\in [0,1),
\end{align}
where $h$ solves the ODE in \eqref{hODE}. Based on $h$'s properties, the function $g$ satisfies 
$$
g'(0) = g(1^-) =0, \quad g'(y) <0,\quad y \in (0,1).
$$
The proof of the next result is given at the end of the next section. {\color{black} Among other things, the below parameter restriction on $A$  and $\delta<0$ give
\begin{align}\label{Dfinite}
\E\Big[ \int_0^\infty e^{-\beta_1t } D_t^{1-\gamma} dt \Big] =\frac{D_0^{1-\gamma} }{\beta_1 -\frac12(1-\gamma) \left(2 \mu_D -\gamma \sigma_D^2\right)}<\infty,
\end{align}
where the geometric Brownian motion $D$ is from \eqref{dD}.}
\begin{theorem}\label{Main} Let $\gamma \in (0,1)$, $\sigma^2_D>0$, and assume the constants 
\begin{align}\label{Def_delta}
\delta := \frac{2(\beta_2-\beta_1)}{\sigma_D^2}
\end{align}
and 
\begin{align}\label{Def_A}
A:= \frac{2 \beta_2+\sigma_D^2 -(1-\gamma) (2 \mu_D -\gamma  \sigma_D^2)}{\sigma_D^2}
\end{align}
satisfy $A \in (1+\delta - \frac{2\delta}{\gamma},\infty)$ and $\delta \in (-\gamma,0)$. For $\theta^{(0)}_{2,0-}\in \big(0,{\color{black}g(0)D_0}\big)$, let $Y_0 \in (0,1)$ {\color{black} be the unique solution of} $g(Y_0) D_0(1-Y_0)^\gamma=\theta^{(0)}_{2,0-}$. Then, there exists a Radner equilibrium in which 
the  equilibrium interest rate process is $r_t=r(Y_t)\in \sL_{\text{loc}}^1$ and the equilibrium market price of risk process is $\kappa_t = \kappa(Y_t)\in \sL_{\text{loc}}^2$ for the deterministic functions
\begin{align}\label{randkappa}
\begin{split}
r(y) & := \beta_2 + y(\beta_1-\beta_2) +\gamma  \mu_D -\frac{1}{2} \gamma (\gamma +1) \sigma_D^2-\frac{\gamma  (\gamma +1) \sigma_D^2(1-y)}{2  y h(y)^2},\\
\kappa(y) &:= \gamma  \sigma_D  \left(\frac{1-y}{y h(y)}+1\right),
\end{split}
\end{align}
for $y\in(0,1)$ and the equilibrium consumption-rate processes for \eqref{primalval} and \eqref{primalval1} are 
\begin{align}\label{inhome15}
\hat{c}_{1,t}:=D_tY_t,\quad\hat{c}_{2,t}:=D_t(1-Y_t),\quad t\ge0,
\end{align}
where the state-process $Y$ is as in \eqref{inhomeY}.
\end{theorem}
\noindent Based on \eqref{inhome15}, we say that $Y$ is the equilibrium consumption-share process of trader 1.

\subsection{Survival analysis}\label{survival}

As in Kogan, Ross, Wang, and Westerfield (2006)  and Yan (2008), we use the consumption share process \eqref{inhomeY} to determine if a trader survives in the long run. 
\begin{definition}
 Trader 1, respectively Trader 2, becomes extinct iff
$$
\lim_{t\to\infty} Y_t = 0,\quad \text{respectively }\lim_{t\to\infty} (1-Y_t) = 0,\quad \text{almost surely}.
$$
Otherwise, Trader 1, respectively Trader 2, is said to survive.
$\endproof$
\end{definition}
\noindent Based on this definition, even if one of the consumption shares converges to zero in the sense $\lim_{t\to\infty} Y_t \in \{0,1\}$, the corresponding equilibrium consumption-rate process $\hat c_1$ or  $\hat c_2$  in \eqref{inhome15} may or may not converge to zero. This is because the geometric Brownian motion in \eqref{dD} satisfies $\lim_{t\to\infty} D_t = \infty$ whenever $\mu_D>\frac12\sigma_D^2$.

For $\beta_1=\beta_2$, Lemmas 3.2 and 3.3 in Guasoni, Larsen, and Leoni (2025) ensure that the scale function $s$ defined in \eqref{scalefct} below satisfies $s(0^+) =-\infty$ and $s(1^-) <\infty$ and ensure that both boundaries $y=0$ and $y=1$ are not attainable. Therefore, for  $\beta_1=\beta_2$, Proposition 5.5.22(c) in Karatzas and Shreve (1998) gives $\lim_{t\to\infty}Y_t =1$ almost surely. 
This property also holds in Basak and Cuoco (1998)  where $\gamma =1$ and $h(y) = 1$ for all $y \in [0,1]$. Consequently, in these two models, the restricted trader becomes extinct. In contrast, the next result shows that 
when  $\beta_2 <\beta_1$,  both traders can survive.

\begin{lemma}\label{lemma_survival} Assume the setting of Theorem \ref{Main}. {\color{black} When $\delta \in (-\gamma, -\gamma^2]$, both traders survive. When $\delta \in (-\gamma^2,0)$, Trader 1 survives and Trader 2 becomes extinct.}
\end{lemma}
\proof {\color{black} This follows from Lemma \ref{MainLemma}(2).}
$\endproof$

We conclude this section by showing that the model in Prieto (2013) can also produce surviving traders. In Prieto (2013), the restricted trader has a log-utility function, the unrestricted trader has a  power-utility function. As discussed in the introduction, when the restricted trader has a log-utility function, the optimization problem {\color{black}
\begin{align}\label{primalval1_log}
&\sup_{ c_2\in \sA_2}\E\left[\int_0^\infty e^{-\beta_2t}\log(c_{2,t})dt\right]
\end{align}
 is explicitly solvable with optimizer
\begin{align}\label{primalval1_log2}
\hat c_{2,t} := \hat X_{2,t}\beta_2,\quad d\hat X_{2,t} = \hat X_{2,t}( r_t - \beta_2)dt.
\end{align}
In \eqref{primalval1_log}, we adopt the convention that the log objective equals $-\infty$ whenever the consumption rate equals zero  on a set of positive measure.

In the remainder of this subsection, we replace trader 2’s objective \eqref{primalval1} with the logarithmic
objective \eqref{primalval1_log}, and replace item (ii) of Definition \ref{def:eq}  accordingly. The next theorem's assumption $\theta^{(0)}_{2,0-}\in \big(0,{\frac{D_0}{\beta_2}}\big)$ was already used in Basak and Cuoco (1998). 

\begin{theorem}\label{thm:Prieto}Assume $2\beta_1 > (1-\gamma)(2\mu_D-\gamma\sigma_D^2)$, $\theta^{(0)}_{2,0-}\in \big(0,{\frac{D_0}{\beta_2}}\big)$, and set $Y_0 :=1-\frac{\beta_2\theta^{(0)}_{2,0-}}{D_0}$.

\begin{enumerate}
 
\item (Prieto) There exists a Radner 
equilibrium in which \eqref{inhome15} holds for the state process \eqref{inhomeY} with coefficient functions
\begin{align*}
\mu_Y(y)&:=\frac{(y-1) \left(2 y^2 (-\beta_1+\beta_2-\gamma  \mu_D+\mu_D)+\sigma_D^2 \left(\gamma ^2+\gamma -2 y^2+2 \gamma  (y-1) y\right)\right)}{2 y ((\gamma -1) y-\gamma )},\\
\sigma_Y(y)&:= \sigma_D(1-y),\quad y \in (0,1). 
\end{align*}

\item If the inequality 
\begin{align}\label{prieto1}
\eta:=\frac{\beta_2-\beta_1+(1-\gamma)  \mu_D}{\sigma_D^2}-\frac{1}{2} (1-\gamma ) (\gamma +2)\le-\frac12.
\end{align}
holds, both traders survive in the long run.

\end{enumerate}

\end{theorem}

}

\section{Proofs}

This section adjusts the proofs in Guasoni, Larsen, and Leoni (2025) to accommodate different time-preference coefficients $\beta_1$ and $\beta_2$. Mathematically speaking, when $\beta_1\neq\beta_2$, the governing ODE \eqref{hODE} has a new cubic term proportional to $\beta_2-\beta_1$.

\subsection{Auxiliary ODE analysis}
In this subsection, we consider the ODE
 \begin{align}\label{fODE}
\begin{cases}
f'(y)  =a_0(y)+ \frac{a_1(y)}{1-y}f(y)+\frac{a_3}{1-y}f(y)^2 + \frac{\delta y}{1-y}f(y)^2\Big ( 1- \frac{f(y)}\gamma\Big),\quad y\in(y_0,1), \\
f(y_0) = f_0,
\end{cases}
\end{align}
for constants $y_0,f_0,a_3,\delta,\gamma \in\R$ and functions $a_0$ and $a_1$ from \eqref{a0a1}.
\begin{theorem}\label{thm:f} Let $\gamma \in (0,1)$,  $\delta \in (-\gamma,0)$, and $a_3 \in [-1,\delta\frac{1-\gamma}{\gamma}-\gamma)$. 
\begin{enumerate}
\item For $y_0:=0$ and $f_0 := \gamma$, there exists $f\in \sC([0,1])\cap \sC^1([0,1))$ such that \eqref{fODE} holds with $\gamma \le f(y)<1$ for all $y\in[0,1]$ and
\begin{align}\label{f1}
f(1) = \frac{\gamma}{2\delta}\Big(a_3 +\delta +\sqrt{(a_3+\delta)^2 + 4\delta}\Big) <1.
\end{align}
\item For $y_0\in (0,1)$ and $f_0 \in {\color{black}(0, 1)}$, there exists $f\in \sC([y_0,1])\cap \sC^1([y_0,1))$ such that \eqref{fODE} and \eqref{f1} hold and $f(y)<1$ for all $y\in[y_0,1]$. 
\end{enumerate}
\end{theorem}
\proof  The second part is easier to prove because there is no singularity at $y_0\in (0,1)$, and so for brevity we only prove the first part.

\noindent{\bf Step 1/4:} This step ensures that all coefficient restrictions are internally consistent.  
\noindent (i) To see that the interval $[-1,\delta\frac{1-\gamma}{\gamma}-\gamma]$ is a non-trivial subinterval of $[-1,0)$, we use  $\gamma \in (0,1)$ and  $\delta \in (-\gamma,0)$ to see 
$$
 -1 < \delta\frac{1-\gamma}{\gamma}-\gamma <0.
$$
\noindent (ii) To see that the term inside the square-root in \eqref{f1} is positive, we use $\delta<0$ to see that the function 
\begin{align}\label{f10}
[-1,0) \ni a_3 \to (a_3+\delta)^2
\end{align}
is decreasing. This gives
\begin{align}\label{f100}
(a_3+\delta)^2 +4\delta {\color{black} \,  > \,} \Big(\delta\frac{1-\gamma}{\gamma}-\gamma+\delta\Big)^2  +4\delta = \frac{\left(\gamma ^2+\delta \right)^2}{\gamma ^2} {\color{black} \,\ge\,}0.
\end{align}
\noindent (iii) To see that the function 
\begin{align}\label{f1a}
[-1,\delta\frac{1-\gamma}{\gamma}-\gamma]\ni a_3\to a_3 +\delta +\sqrt{(a_3+\delta)^2 + 4\delta}
\end{align}
is decreasing, we compute the derivative
$$
\frac{\partial}{\partial a_3}\Big(a_3 +\delta +\sqrt{(a_3+\delta)^2 + 4\delta}\Big) = 1+\frac{a_3 +\delta }{\sqrt{(a_3+\delta)^2 + 4\delta}}.
$$
Therefore, the function \eqref{f1a} is decreasing if and only if it is negative. In other words, we need
\begin{align}\label{f1b}
-(a_3 +\delta )\ge \sqrt{(a_3+\delta)^2 + 4\delta}.
\end{align}
Because both sides of \eqref{f1b} are positive, we can square and use $\delta <0$ to see that  \eqref{f1b} holds.

\noindent (iv) To see the upper bound in \eqref{f1}, we evaluate the function in
\eqref{f1a} at $a_3:=\delta\frac{1-\gamma}{\gamma}-\gamma$ to see
$$
 \frac{\gamma}{2\delta}\Big(a_3 +\delta +\sqrt{(a_3+\delta)^2 + 4\delta}\Big)< \frac{|\gamma ^2+\delta |-\gamma^2 +\delta }{2 \delta },\quad a_3 \in (-\infty,\delta\frac{1-\gamma}{\gamma}-\gamma).
$$
By splitting into two cases $\gamma ^2+\delta\ge0$ and $\gamma ^2+\delta<0$, we get 
$$
\frac{|\gamma ^2+\delta |-\gamma^2 +\delta }{2 \delta }=
 \begin{cases}
 -\frac{\gamma^2}{\delta},\quad \delta \in (-\gamma,-\gamma^2]\\
1,\quad \delta \in (-\gamma^2,0)
\end{cases}
\le 1.
$$

\noindent {\bf Step 2/4:} Relative to Theorem 2.4 in Guasoni, Larsen, and Leoni (2025), when $\delta \neq 0$, the term 
\begin{align}\label{cubic1}
 \frac{\delta y}{1-y}f(y)^2\Big ( 1- \frac{f(y)}\gamma\Big),\quad y\in [0,1),
 \end{align} 
 in  \eqref{fODE} is new. However, the local existence and comparison results in Theorems 2.2 and 2.3 in  Guasoni, Larsen, and {\color{black}Leoni }(2025) continue to hold for the ODE in \eqref{fODE}. This is because the  singularity in \eqref{cubic1} is at $y=1$, whereas Theorems 2.2 and 2.3 in Guasoni, Larsen, and Leoni (2025) are local around the initial point $y=0$.

\noindent {\bf Step 3/4:} For $a_3:=-1$, we see that the constant $f(y) := \gamma$ solves  \eqref{fODE}. Therefore, for $a_3\ge -1$, the comparison principle ensures that all local solutions of  \eqref{fODE} are lower bounded by $\gamma$.

\noindent {\color{black}\bf Step 4/4:} This step proves that a global solution of  \eqref{fODE} exists. To this end, we let $f(y)>0$ be a local solution for $y\in[0,y^*)$ where $y^* \in (0,1]$ gives $f$'s maximal interval of existence. 

First, 
to see that $f(y) <1$ for $y\in[0,y^*)$, we argue by contradiction and assume there exists $y_1\in [0,y^*)$ such that $f(y_1) \ge 1$. Then, because $f(0) =\gamma <1$,  there exists $y_0 \in (0,y^*)$ with $f(y_0) =1$ and $f'(y_0)\ge0$. The ODE in  \eqref{fODE}  gives the contradiction
 \begin{align}\label{fODE4}
 \begin{split}
 0 &\le a_0(y_0)+ \frac{a_1(y_0)}{1-y_0}+\frac{a_3}{1-y_0} + \frac{\delta y_0}{1-y_0}\Big ( 1- \frac{1}\gamma\Big)\\
& \le a_0(y_0)+ \frac{a_1(y_0)}{1-y_0}+\frac{\delta\frac{1-\gamma}{\gamma}-\gamma}{1-y_0} + \frac{\delta y_0}{1-y_0}\Big ( 1- \frac{1}\gamma\Big)\\
&=\frac{(\gamma -1) \left(\gamma ^2+\gamma -\delta  y_0\right)}{\gamma  y_0} \\
&<0.
 \end{split}
\end{align}

Second, to see that $\lim_{y\uparrow y^*} f(y)$ exists, it suffices to rule out finite oscillations because of the previous boundedness property. For $y^* \in (0,1)$, there is no singularity in \eqref{fODE} and a standard Lipschitz argument rules out oscillations. To rule out  finite oscillations for $y^*=1$, we let $(y_n)_{n\in\N} \subset (0,1)$ converge to $y^*=1$ such that $f'(y_n) =0$. Because $f$ is bounded, by using a subsequence if necessary, we can assume 
$$
l := \lim_{n\to\infty}f(y_n)  
$$
exists in $[\gamma, 1]$. The proof is concluded by showing that there is only one possible value for $l$. Multiplying $1-y$ on both sides in \eqref{fODE} and replacing $y$ with $y_n$ give the limit
 \begin{align}\label{fODE5}
 \begin{split}
0&=  \gamma l+a_3l^2 + \delta l^2\Big ( 1- \frac{l}\gamma\Big),\\
 \end{split}
\end{align}
The cubic polynomial in \eqref{fODE5} has the 3 roots $l\in \Big\{0, \frac{\gamma}{2\delta}\Big(a_3 +\delta \pm \sqrt{(a_3+\delta)^2 + 4\delta}\Big)\Big\}$. Because  $\gamma \le f \le 1$, we have $l\in [\gamma,1]$ and so it suffices to prove 
 \begin{align}\label{fODE10}
\frac{\gamma}{2\delta}\Big(a_3 +\delta - \sqrt{(a_3+\delta)^2 + 4\delta}\Big)>1
\end{align}
 From \eqref{f100} we have
 \begin{align}
-\sqrt{(a_3+\delta)^2 +4\delta} \le - \frac{|\gamma ^2+\delta|}{\gamma} .
\end{align}
By splitting into two cases $\gamma ^2+\delta\ge0$ and $\gamma ^2+\delta<0$, we get 
$$
a_3 + \delta - \frac{|\gamma ^2+\delta|}{\gamma} < \frac{\delta}{\gamma} -\gamma - \frac{|\gamma ^2+\delta|}{\gamma} =
 \begin{cases}
\frac{2\delta}{\gamma},\quad \delta \in (-\gamma,-\gamma^2]\\
-2\gamma ,\quad \delta \in (-\gamma^2,0)
\end{cases}
\le \frac{2\delta}{\gamma},
$$
which shows \eqref{fODE10}.

$\endproof$

\subsection{Governing ODE analysis}
 Relative to Guasoni, Larsen, and Leoni (2025), when $\delta \neq 0$, the term 
\begin{align}\label{cubic11}
 \frac{\delta y}{1-y}h(y)^2\Big ( 1- \frac{h(y)}\gamma\Big),\quad y\in [0,1),
 \end{align} 
 in  \eqref{hODE} is new. As in the Step 2 in the proof of Theorem \ref{thm:f}, for $\xi \ge0$, the existence of a local solution $h_\xi(y)>0$ of \eqref{hODE} for $y$ near 0 with $h(0)=\gamma$ follows as in Theorem 2.5 in Guasoni, Larsen, and Leoni (2025). This is  because the cubic term \eqref{cubic11} has no singularity at $y=0$.

Uniqueness of local solutions of \eqref{hODE} with $h(0) =\gamma$ follows from the following Lipschitz estimates.  For $\xi \ge0$, we define
\begin{align}\label{F_xi}
\begin{split}
y_\xi &:= \inf\{y>0: h_\xi(y) =1\} \land 1,\\
F_\xi(y) &:= \frac{\xi}{\sigma_D^2} \exp\Big\{ \int_0^y\frac{h_\xi(q)-1}{1-q}dq\Big\},\quad y\in [0,y_\xi].
\end{split}
\end{align}

\begin{lemma}\label{Lem_Lip}  Let $\gamma \in (0,1)$, $\sigma_D^2>0$, $A>1$, $\delta <0$, $y_0 \in (0,1)$, and $\bar\xi>0$. Then, there exist constants $M_1>0$ and $M_2>0$ such that
\begin{align}
|h_{\xi_1}(y)-h_{\xi_2}(y)|\leq M_1 y|\xi_1-\xi_2|,\quad y \in [0, y_{\xi_1}\land  y_{\xi_2} \land y_0],\quad \xi_1,\xi_2\in [0,\bar\xi], \label{Lipp1}\\
|F_{\xi_1}(y)-F_{\xi_2}(y)|\leq M_2 |\xi_1-\xi_2|,\quad y \in [0, y_{\xi_1}\land  y_{\xi_2} \land y_0],\quad \xi_1,\xi_2\in [0,\bar\xi].\label{Lipp2}
\end{align}

\end{lemma}

\proof  For  $\xi_1,\xi_2\in [0,\bar\xi]$,  we let $h_{1}, h_{2}\in \sC^1([0,y_{\xi_i}))$ be the corresponding local solutions of  \eqref{hODE}. We rewrite \eqref{hODE} as
\begin{align}\label{hODE_new1}
\begin{split}
&h_{i}^{\prime}(y)+\frac{1+\gamma}{y}\big(h_{i}(y)-\gamma\big)\\
&=\frac{\gamma}{1-y}h_{i}(y)+\frac{F_{\xi_i}(y)-A}{1-y}h_{i}(y)^2 + \frac{\delta y}{1-y}h_i(y)^2\Big ( 1- \frac{h_i(y)}\gamma\Big),
\end{split}
\end{align}
for $y\in(0, y_{\xi_i})$. Subtracting and multiplying by $y^{1+\gamma}$ give us
\begin{align*}
&  y^{1+\gamma}\big(h_{1}^{\prime}(y)-h_{2}^{\prime}(y)\big)+y^{\gamma}(1+\gamma
)\big(h_{1}(y)-h_{2}(y)\big)\\
&=y^{1+\gamma}\frac{\gamma}{1-y}\big(h_{1}(y)-h_{2}(y)\big)\\
&  +y^{1+\gamma}\frac{F_{\xi_1}(y)-F_{\xi_2}(y)}{1-y}
h_{1}(y)^2+y^{1+\gamma}\frac{F_{\xi_2}(y)-A}{1-y}\big(h_{1}(y)^2-h_{2}(y)^2\big)\\
&+ y^{1+\gamma}\frac{\delta y}{1-y}\bigg(h_1(y)^2\Big ( 1- \frac{h_1(y)}\gamma\Big)-h_2(y)^2\Big ( 1- \frac{h_2(y)}\gamma\Big)\bigg), \quad y <  y_{\xi_1}\land  y_{\xi_2}.
\end{align*}
Because $0\le h_i \le 1$, we have the bounds
\begin{align*}
&\big|h_{1}(y)^2-h_{2}(y)^2\big|  = \big(h_{1}(y)+h_{2}(y)\big)\big|h_{1}(y)-h_{2}(y)\big| \le 2\big|h_{1}(y)-h_{2}(y)\big|,
\end{align*}
and 
\begin{align*}
&\bigg|h_1(y)^2\Big ( 1- \frac{h_1(y)}\gamma\Big)-h_2(y)^2\Big ( 1- \frac{h_2(y)}\gamma\Big)\bigg|\\
&=
\big|h_1(y)-h_2(y)\big| \left|h_1(y)+h_2(y)-\frac{h_1(y) h_2(y)+h_1(y)^2+h_2(y)^2}{\gamma }\right| \\
&\le \big|h_1(y)-h_2(y)\big| \big(2+\frac3\gamma \big).
\end{align*}
These bounds allow us to use Gronwall's inequality to derive the bounds \eqref{Lipp1}-\eqref{Lipp2}. Because the arguments are identical to those in the proof of Lemma 2.8 in Guasoni, Larsen, and Leoni (2025), we omit the details.

$\endproof$

\noindent \emph{Proof of Lemma \ref{MainLemma}(1):}  The following proof adjusts the proof of Theorem 1.1 in Guasoni, Larsen, and Leoni (2025) to include {\color{black}the new term} \eqref{cubic11}.

\noindent {\bf Step 1/7:} {\color{black}Because $\delta >-\gamma$, the interval $\big((A-1)\sigma_D^2, (A+\delta \frac{1-\gamma}{\gamma} -\gamma)\sigma_D^2\big)$ is not empty.} For $\xi \in ({\color{black}(A-1)\sigma_D^2}, (A+\delta \frac{1-\gamma}{\gamma} -\gamma)\sigma_D^2)$, this step ensures that a global solution to  \eqref{hODE} and $h(0) =\gamma$ exists. We let $f$ be the solution of \eqref{fODE} produced by Theorem \ref{fODE} for
$$
a_3:= \frac{\xi}{\sigma_D^2} - A < \delta \frac{1-\gamma}{\gamma} -\gamma,\quad {\color{black}a_3\ge-1}.
$$
We define
$$
y_1:= \inf\{y>0: h(y)=1\} \in (0,1] \cup \{\infty\}.
$$
To see $y_1 = \infty$, we assume to the contrary that $y_1 \in (0,1]$. Continuity of $h$ gives $h(y_1) =1$. However,  \eqref{a0a1} gives $a_2(h,y)\le  \frac{\xi}{\sigma_D^2}-A = a_3$ for $y \in [0,y_1]$ and the comparison principle produces $h\le f<1$.

To rule out finite oscillations at some interior point $y^* \in (0,1)$, we note that there is no singularity in \eqref{fODE} and a standard Lipschitz argument ensures that $\lim_{y\uparrow y^*}h(y)$ exists. To rule out  finite oscillations at $y^*=1$, we note that $h(y)<1$ for $y \in [0,1)$ gives $a_2(h,y)\le  \frac{\xi}{\sigma_D^2}-A = a_3$ for $y \in [0,1)$, hence, the comparison principle gives $h\le f<1$. Therefore,
\begin{align}\label{zerolimitintegral}
0\le \limsup_{y\uparrow 1} \exp\left\{\int_0^y\frac{h(q)-1}{1-q}dq\right\} \le \limsup_{y\uparrow 1} \exp\left\{\int_0^y\frac{f(q)-1}{1-q}dq\right\} =0,
\end{align}
where the last equality uses $f(1) <1$. All in all, $ \lim_{y\uparrow 1} \exp\big\{\int_0^y\frac{h(q)-1}{1-q}dq\big\} =0$. To see that $ \lim_{y\uparrow 1} h(y)$ exists, we proceed as in Step 4/4 of the proof of Theorem \ref{thm:f} and let $(y_n)_{n\in\N} \subset (0,1)$ converge to $y^*=1$ such that $h'(y_n) =0$. Because $h$ is bounded, by using a subsequence if necessary, we can assume $l:=\lim_{n\to\infty} h(y_n) \in [0,1)$ exists and solves the analogue of  
\eqref{fODE5} given by
\begin{align}\label{3degreepol}
 \begin{split}
0&=  \gamma l-Al^2 + \delta l^2\Big ( 1- \frac{l}\gamma\Big).
 \end{split}
\end{align}
Similar to Step 4/4 of the proof of Theorem \ref{thm:f}, the cubic equation  \eqref{3degreepol} has exactly one solution in $(0,1)$, which is given by 
\begin{align}\label{h1}
l = \frac{\gamma}{2\delta}\Big( \delta - A + \sqrt{(A-\delta)^2 + 4\delta}\Big)\le \gamma.
\end{align}
The upper bound in \eqref{h1} comes from $A\ge 1$. To rule out $l=0$ as a possible limit, we argue by contradiction to see
 \begin{align*}
0  &= \lim_{n\to\infty} \bigg(a_0(y_n)+ \frac{h(y_n)}{1-y_n}\Big(a_1(y_n)+a_2(h,y_n)h(y_n) + \delta y_n h(y_n)\big ( 1- \frac{h(y_n)}\gamma\big)\Big)\bigg)\\
 &= \gamma(1+\gamma) + \lim_{n\to\infty} \frac{h(y_n)}{1-y_n}\gamma.
\end{align*}
This gives a contradiction because $h\ge0$. All in all,  $h$ cannot oscillate and $ \lim_{y\uparrow 1} h(y)$ exists and equals $l$ in \eqref{h1}.\ \\

\noindent {\bf Step 2/7:} For $y_0 \in (0,1)$, this step proves that $\lim_{\xi \uparrow \infty} h(y_0) = \infty$. {\color{black} Assume the contrary. We decompose the ODE \eqref{hODE} as
\[
    \frac{\delta y}{1-y}h(y)^2
    \left(
        1-\frac{h(y)}{\gamma}
    \right)
    =
    \frac{\delta y}{1-y}h(y)^2
    -
    \frac{\delta y}{\gamma(1-y)}h(y)^3.
\]
Since $\delta<0$, $y\in[0,1)$, and $h(y)\geq0$,  the cubic term satisfies
\[
    -\frac{\delta y}{\gamma(1-y)}h(y)^3\geq0.
\]
Therefore,
 \begin{align}\label{hODE35}
 \begin{split}
h'(y)
&=
a_0(y)
+
\frac{a_1(y)}{1-y}h(y)
+
\frac{a_2(h,y)+\delta y}{1-y}h(y)^2
-
\frac{\delta y}{\gamma(1-y)}h(y)^3
\\
&\geq
a_0(y)
+
\frac{a_1(y)}{1-y}h(y)
+
\frac{a_2(h,y)+\delta y}{1-y}h(y)^2.
\end{split}
\end{align}
}{\color{black} To produce a contradiction, we define the function $g_\xi(y) := h_\xi(y) (1-y)^\gamma y^{1+\gamma}$, which satisfies 
\begin{align}\label{compa1}
\begin{split}
g'_\xi(y) &\ge a_0(y)(1-y)^\gamma y^{1+\gamma}+  \tfrac{a_2(h,y)+\delta y}{\big((1-y)y\big)^{1+\gamma}}g_\xi(y)^2\\
&\ge \gamma(1+\gamma)(1-y)^\gamma y^\gamma+  \tfrac{\frac{\xi}{\sigma_D^2} (1-y)-A+\delta y}{\big((1-y)y\big)^{1+\gamma}}g_\xi(y)^2
,\quad y\in(0,y_0).
\end{split}
\end{align}
By choosing $\xi>0$ such that $\frac{\xi}{\sigma_D^2} (1-y_0)-A+\delta y_0>0$, we can use the proof of Lemma 2.7 in Guasoni, Larsen, and Leoni (2025) to construct  a lower bound for $g_\xi$, which explodes at some $y \in (0,y_0]$.} \ \\

\noindent {\bf Step 3/7:} As in Guasoni, Larsen, and Leoni (2025), we define the subset $\Xi$ of $(0,\infty)$ by
$$
\Xi:= \big\{ \xi >0 :  \text{$h\in \sC([0,1])\cap \sC^1([0,1))$ solves \eqref{hODE} with }  h(0) = \gamma \text{ and }h(1) \le  \gamma\big\}.
$$
{\color{black} Step~1 shows $\xi \in \big((A-1)\sigma_D^2, (A+\delta \frac{1-\gamma}{\gamma} -\gamma)\sigma_D^2\big) \subset \Xi$} produces a solution $h$ with $h<1$. This step generalizes this property to all $\xi \in \Xi$. The proof is similar to the proof of Lemma 2.10(2) in Guasoni, Larsen, and Leoni (2025).  We assume for the sake of contradiction that there exists
{\color{black}$y'_0\in[0,1]$ with $h(y'_0)\geq1$}.
Since {\color{black}  $h\in \sC([0,1])$} and $h(0)=\gamma<1$ and $h(1)\leq\gamma<1$, the maximum of $h$
is attained at some {\color{black}$y_0\in(0,1)$ with  $h(y_0)\geq1$}. Because $h\in \sC^2((0,1))$, we have
$$
h'(y_0) = 0,\quad h''(y_0)\le 0.
$$
Inserting $h'(y_0)=0$ into \eqref{hODE} produces
\begin{align}\label{a2_113}
a_2(h,y_0)= -\frac{(1-y_0) a_0(y_0)}{h(y_0)^2}-\frac{a_1(y_0)}{h(y_0)}-\frac{\delta  y_0\big(\gamma-h(y_0)\big)}{\gamma }.
\end{align}
By using $h'(y_0) = 0$ and \eqref{a2_113} when computing the derivative of \eqref{hODE}, we get
\begin{align}\label{hdouble}
\begin{split}
h''(y_0) & = \frac{\big(h(y_0)-\gamma\big) \big(\gamma ^2+\gamma -\delta  y_0^2 h(y_0)^2\big)}{\gamma  (1-y_0) y_0^2} + \frac{h(y_0)^2}{1-y_0}\frac{\partial }{\partial y} a_2(h,y_0)\\
&\ge \frac{(\gamma +1) \big(h(y_0)-\gamma \big)}{(1-y_0) y_0^2}.
\end{split}
\end{align}
The inequality in \eqref{hdouble} comes from $\delta <0$ and 
$$
\frac{\partial }{\partial y} a_2(h,y) =  \frac{\xi}{\sigma_D^2}\exp\Big\{\int_0^y \frac{h(q)-1}{1-q}dq\Big\}\frac{h(y)-1}{1-y},
$$
which is non-negative at $y=y_0$ because we have assumed $h(y_0) \ge 1$. The second line in \eqref{hdouble} is strictly positive, which contradicts $h''(y_0) \le 0$. \\

\noindent {\bf Step 4/7:} Step 1 ensures that $\Xi \neq \emptyset$ and Step 2 ensures that $\Xi$ is a bounded subset of $(0,\infty)$. Consequently, $\xi_0 := \sup \Xi \in (0,\infty)$. This step proves $\xi_0 \notin \Xi$. We argue by contradiction and assume $\xi_0 \in \Xi$ and let $h_{\xi_0}$ denote the corresponding solution to  \eqref{hODE}. Next, {\color{black}we will use Lemma \ref{Lem_Lip} to construct some $\xi \in (\xi_0, \xi_0+1)$ with $\xi \in \Xi$}. 
The assumption $\xi_0 \in \Xi$ gives $h_{\xi_0}(1) \le \gamma<1$ and, similarly to \eqref{zerolimitintegral}, we have $
\lim_{y\uparrow 1} F_{\xi_0}(y)  = 0$. Therefore, because $A>1$, we can find $y_0 \in (0,1)$ such that
\begin{align}\label{Fbound}
\forall y \in (y_0,1): F_{\xi_0}(y)  <\frac {A-1}{2}.
\end{align}
 From \eqref{Lipp1}, there exists a constant $M_1$ such that
\begin{align}
\forall \xi\in (\xi_0,\xi_0+1)\;\forall y \in [0, y_{\xi} \land y_0]:\quad |h_{\xi}(y)-h_{\xi_0}(y)|\leq M_1 |\xi-\xi_0|.\label{Lipp1a}
\end{align}
Because  $h_{\xi_0} <1$, we can find $\xi  \in (\xi_0,\xi_0+1)$ such that
\begin{align}\label{xi1}
h_\xi(y) \le |h_{\xi}(y)-h_{\xi_0}(y)| + h_{\xi_0}(y) \le M_1 |\xi-\xi_0| + \sup_{y\in[0,1]} h_{\xi_0}(y) <1,
\end{align}
for $y \le y_{\xi}\land y_0 = y_0$.  From \eqref{Lipp2},  there exists a constant $M_2$ such that
\begin{align}\label{F1}
\forall \xi\in (\xi_0,\xi_0+1)\;\forall y \in [0, y_{\xi} \land y_0]:\quad |F_{\xi}(y)-F_{\xi_0}(y)|\leq M_2 |\xi-\xi_0|.
\end{align}
Let $\xi \in (\xi_0,\xi_0+1)$ satisfy \eqref{xi1}. Because $y_{\xi}> y_0$, the inequality \eqref{F1} gives 
\begin{align}\label{xi2}
F_\xi(y_0) \le |F_{\xi}(y_0)-F_{\xi_0}(y_0)| + F_{\xi_0}(y_0) \le M_2 |\xi-\xi_0| + \frac{A-1}{2},
\end{align}
where the last inequality uses \eqref{Fbound}. Because $M_2$ does not depend on $\xi$, we can {\color{black} choose $\xi>\xi_0$ sufficiently close to $\xi_0$} so that \eqref{xi2} ensures  $F_\xi(y_0) <A-1$.  Because $F_{\xi}(y)$ is decreasing in $y \in [0,y_\xi]$, we can use the comparison principle to see $h_\xi (y) \le f(y) <1$ for $y \in [y_0,y_\xi]$ for $f$ given by Theorem \ref{thm:f}(2) with $a_3:=-1$ and $f(y_0) = h_\xi(y_0)$. Therefore,  $y_\xi =1$ and  $\xi > \xi_0$ {\color{black}and  $h_\xi (1) \le f(1) = \gamma <1$. Hence, \(\xi\in\Xi\), which contradicts
\(\xi>\xi_0=\sup\Xi\).}

\ \\
\noindent {\bf Step 5/7: }{\color{black}Let $(\xi_n)_{n\in\mathbb N}\subset\Xi$ be an increasing
sequence with $\xi_n\uparrow\xi_0:=\sup\Xi$. The comparison argument
in Theorem~2.5(2) of Guasoni, Larsen, and Leoni (2025) is unchanged
because the cubic term is common to both equations. Hence, 
$h_{\xi_n}\leq h_{\xi_{n+1}}$ and the pointwise limit
$$
h_{\xi_0}(y):= \lim_{n\to\infty} h_{\xi_n}(y),\quad y\in [0,1),
$$
exists and satisfies $0\le h_{\xi_0}\le 1$.

For $n\in\N$, the ODE in \eqref{hODE} produces the integral representation
\begin{align*}
&y^{1+\gamma}(1-y)^\gamma h_{\xi_n}(y) \\
&= \int_0^yq^{1+\gamma}(1-q)^\gamma \bigg(a_0(q) +\frac{a_2(h_{\xi_n},q)}{1-q}h_{\xi_n}(q)^2  
 + \frac{\delta q}{1-q}h_{\xi_n}(q)^2\Big ( 1- \frac{h_{\xi_n}(q)}\gamma\Big)\bigg)dq,
\end{align*}
for $y \in [0,1)$.  To use the Dominated Convergence Theorem, we use \(0\leq h_{\xi_n}\leq1\) and \(\xi_n\leq\xi_0\), to see
\[
    0<
    \exp\left\{
        \int_0^q
        \frac{h_{\xi_n}(r)-1}{1-r}\,dr
    \right\}
    \leq1.
\]
Consequently,
\[
    \left|a_2(h_{\xi_n},q)\right|
    \leq
    A+\frac{\xi_0}{\sigma_D^2},
    \qquad n\in\mathbb N,\quad q\in[0,1).
\]
For the other term in the integrand,  we have when $ h_{\xi_n}(q) \ge \gamma$
$$
0\le \delta h_{\xi_n}(q)^2 \left(1-\frac{h_{\xi_n}(q)}{\gamma}\right)\le 
\delta \left(1-\frac{1}{\gamma}\right),
$$
and when $ h_{\xi_n}(q) < \gamma$ we have 
$$
0\ge \delta h_{\xi_n}(q)^2 \left(1-\frac{h_{\xi_n}(q)}{\gamma}\right)\ge 
\delta.
$$
Therefore, the Dominated Convergence Theorem ensures that $h_{\xi_0}$ satisfies the same integral equation and so satisfies the ODE \eqref{hODE}.}

A small adjustment to the endpoint argument from Step~1
gives that $h_{\xi_0}(1):=\lim_{y\uparrow1}h_{\xi_0}(y)$ exists in $[0,1]$.   Step 4 gives $h_{\xi_0} \notin \Xi$ and so $h_{\xi_0}(1) \in (\gamma,1]$. To see that $h_{\xi_0}(1) \in (\gamma,1)$ is impossible, we can argue as in Step 1. Therefore, the only possibility is $h_{\xi_0}(1)=1$.\\

\noindent {\bf Step 6/7:} The ODE in \eqref{hODE} produces the integral representation
\begin{align}\label{integral_rep}
\begin{split}
&y^{1+\gamma}(1-y)^\gamma h_{\xi_0}(y) \\
&= \int_0^yq^{1+\gamma}(1-q)^\gamma \bigg(a_0(q) +\frac{a_2(h_{\xi_0},q)}{1-q}h_{\xi_0}(q)^2  
 + \frac{\delta q}{1-q}h_{\xi_0}(q)^2\Big ( 1- \frac{h_{\xi_0}(q)}\gamma\Big)\bigg)dq,
\end{split}
\end{align}
for $y \in [0,1)$. Because $ 0 \le h_{\xi_0} \le 1$ and $h_{\xi_0}(0) = \gamma <1$, the following  limit exists
$$
\int_0^1\frac{h_{\xi_0} (q)-1}{1-q}dq := \lim_{y\uparrow 1}\int_0^y\frac{h_{\xi_0} (q)-1}{1-q}dq \in [-\infty,0).
$$
Unlike \eqref{zerolimitintegral}, {\color{black} we shall see that }this limit is finite. To compute the limit, we divide $y^{1+\gamma}(1-y)^\gamma$ on both sides  of \eqref{integral_rep} and use  L'Hopital's rule when passing $y\uparrow 1$ to see
\begin{align}\label{nonzerolimitintegral}
 \exp\left\{\int_0^1\frac{h_{\xi_0} (q)-1}{1-q}dq\right\} = \frac{\sigma_D^2}{\xi_0}\Big( A-\gamma +\delta \frac{1-\gamma}{\gamma}\Big).
\end{align}

To see that $h_{\xi_0} \ge \gamma$, the limit in \eqref{nonzerolimitintegral} and $h_{\xi_0} \le 1$ produce the following bound for
\eqref{a0a1}
$$
a_2(h_{\xi_0},y) \ge \frac{\xi_0}{\sigma_D^2} \exp\left\{\int_0^1\frac{h_{\xi_0} (q)-1}{1-q}dq\right\} -A =\delta \frac{1-\gamma}{\gamma}-\gamma >-1.
$$
The comparison principle gives $h_{\xi_0} \ge f\ge\gamma$ where $f$ is from Theorem \ref{thm:f}  with $a_3:=-1$.

\noindent {\bf Step 7/7:} Because $h_{\xi_0} \le 1$, the difference quotient satisfies
\begin{align}\label{g_derivativeinstep7_7}
k(y) := \frac{1-h_{\xi_0}(y)}{1-y}\ge0,\quad y\in[0,1).
\end{align}
To prove that $\lim_{y\uparrow 1}k(y)$ exists and is identical to  $\lim_{y\uparrow 1}h'_{\xi_0}(y)$, we need a  representation of $k'$ and $k''$. The formula in \eqref{nonzerolimitintegral} allows us to rewrite $F_{\xi_0}$ in \eqref{F_xi} as
$$
F_{\xi_0}(y) = \Big( A-\gamma +\delta \frac{1-\gamma}{\gamma}\Big)\exp\Big\{ \int_y^1k(q)dq\Big\}, \quad y\in [0,1].
$$
Inserting this expression into \eqref{hODE_new1} and using $k'(y) = \frac{k(y) - h'_{\xi_0}(y)}{1-y}$ give
\begin{align}\label{hODE_new100}
\begin{split}
k(y) - (1-y)k'(y) 
&= \frac{1+\gamma}{y}\big(\gamma-h_{\xi_0}(y)\big) +\gamma h_{\xi_0}(y) k(y) \\
&+\Big( A-\gamma +\delta \frac{1-\gamma}{\gamma}\Big)\frac{ \exp\Big\{ \int_y^1k(q)dq\Big\}-1}{1-y}h_{\xi_0}(y)^2 \\
&+ \delta yh_{\xi_0}(y)^2\Big ( \frac{k(y)}\gamma + \frac{1-\gamma}{\gamma y}\Big).
\end{split}
\end{align}

We split the argument into two cases: First, we argue by contradiction to rule out that $k(y)$ increases to infinity as $y\uparrow 1$. {\color{black}The bound $e^x\ge 1+x$ for $x\in\R$}, the Mean-Value Theorem, and the assumed monotonicity of $k(y)\ge0$ for $y$ near $1$ give
\begin{align*}
\exp\Big\{\int_y^1 k(q)dq\Big\} -1\ge {\color{black}\int_y^1k(q)dq} \ge k(y)(1-y).
\end{align*}
Combining this inequality with the ODE in \eqref{hODE_new100} and $k'(y) \ge0$ gives
\begin{align}\label{hODE_new101}
\begin{split}
k(y) 
&\ge\frac{1+\gamma}{y}\big(\gamma-h_{\xi_0}(y)\big) +\gamma h_{\xi_0}(y) k(y) \\
&+\Big( A-\gamma +\delta \frac{1-\gamma}{\gamma}\Big)k(y)h_{\xi_0}(y)^2 
+ \delta yh_{\xi_0}(y)^2\Big ( \frac{k(y)}\gamma + \frac{1-\gamma}{\gamma y}\Big).
\end{split}
\end{align}
Rearranging \eqref{hODE_new101} and passing $y\uparrow 1$ produce the contradiction
$$
1-\gamma^2 +\delta - \frac{\delta}{\gamma} \ge \Big(A - 1-\delta+ 2\frac{\delta}{\gamma}\Big)\lim_{y\uparrow 1}k(y)=\infty.
$$

Second, we consider oscillations. We differentiate \eqref{hODE_new100} at a point $y\in(0,1)$ with $k'(y) = 0$ to get
\begin{align*}
\gamma(1-y)^2y^2k''(y)
&=y k(y) \Big(y h(y) \big(2 A \gamma +e^{\int_y^1k(q)dq} (h(y)-2) (A \gamma -\gamma  (\gamma +\delta )+\delta )-\gamma ^2\\
&+2 \delta  y h(y)-\delta  h(y)-2 \gamma  \delta  y+2 \delta  y\big)+\gamma  (\gamma -(\gamma +2) y+1)\Big)\\
&+(\gamma +1) \gamma ^2+(y-1) y^2 k(y)^2 \left(\gamma ^2+2 \delta  y h(y)\right)\\
&-(\gamma -1) \delta  y^2 h(y)^2-(\gamma +1) \gamma  h(y),\quad y\in(0,1).
\end{align*}
Let $(y_n)_{n\in\N}$ be a sequence of local maxima for $k$ with  $y_n \uparrow 1$ such that
$$
 \forall n\in\N: \quad k'(y_n) =0,\quad k''(y_n) \le 0,\quad \lim_{n\to\infty} k(y_{n}) =  \limsup_{y\uparrow 1} k(y) \in [0,\infty].
$$ 
Because $h_{\xi_0}(1) =1$, we have
$$
\lim_{n\to\infty} (1-y_n) k(y_{n}) = \lim_{n\to\infty}  \big( 1- h_{\xi_0}(y_{n})\big) = 0.
$$
Therefore, the above ODE for $k''$ gives
\begin{align*}
0 &\ge \lim_{n\to\infty}  \gamma (1-y_{n})^2y^2_{n}k''(y_{n}) \\
&=-(1-\gamma ) \left(\gamma ^2+\gamma -\delta \right) +\Big( \gamma  (A-\delta -1)+2 \delta\Big)\lim_{n\to\infty} k(y_{n}).
\end{align*}
This gives the upper bound 
\begin{align}\label{bound1}
\limsup_{y\uparrow 1} k(y) \le \frac{(1-\gamma ) \left(\gamma ^2+\gamma -\delta \right)}{\gamma  (A-\delta -1)+2 \delta}.
\end{align}

Next, let $(y_n)_{n\in\N}$ be a sequence of local minima for $k$ with  $y_n \uparrow 1$ such that
$$
 \forall n\in\N:\quad  k'(y_n) =0,\quad k''(y_n) \ge 0, \quad \lim_{n\to\infty} k(y_{n}) =  \liminf_{y\uparrow 1} k(y) \in [0,\infty].
$$
Then, similarly to \eqref{bound1}, we have the lower bound 
\begin{align}\label{bound2}
\liminf_{n\to\infty}  k(y_n) \ge\frac{(1-\gamma ) \left(\gamma ^2+\gamma -\delta \right)}{\gamma  (A-\delta -1)+2 \delta}.
\end{align}
The two bounds  \eqref{bound1} and \eqref{bound2} produce the limit $\lim_{y\uparrow 1}k(y) = k(1) = \frac{(1-\gamma ) \left(\gamma ^2+\gamma -\delta \right)}{\gamma  (A-\delta -1)+2 \delta}$.

{\color{black}
The formula in \eqref{g_derivativeinstep7_7} and $h_{\xi_0}(1)=1$ show that $h_{\xi_0}$'s
left derivative at $1$ exists and satisfies
\[
h_{\xi_0}'(1)=\lim_{y\uparrow1}k(y)=k(1).
\]
Moreover, L'Hopital's rule gives
\[
\lim_{y\uparrow1}
\frac{\exp\!\left\{\int_y^1 k(q)\,dq\right\}-1}{1-y}
=k(1).
\]
Rewriting the ODE \eqref{hODE} in terms of \(k\) and passing to the limit therefore yields
\[
\lim_{y\uparrow1}h_{\xi_0}'(y)=k(1)=h_{\xi_0}'(1).
\]
Hence, we have \(h_{\xi_0}\in C^1([0,1])\) with $h_{\xi_0}'(1) = \frac{(1-\gamma ) \left(\gamma ^2+\gamma -\delta \right)}{\gamma  (A-\delta -1)+2 \delta}$.
}
$\endproof$

\subsection{Remaining proofs}

\noindent \emph{Proof of Lemma \ref{MainLemma}(2):}   {\bf Step 1/3:} For a constant $a\in (0,1)$, we define the function
\begin{align}\label{rho1}
\rho(y):= \exp\Big\{ -2 \int_a^y \frac{\mu_Y(x)}{\sigma_Y(x)^2}dx\Big\},\quad y \in (0,1),
\end{align}
where the drift $\mu_Y$ and volatility $\sigma_Y$ are defined in \eqref{mgmethod1bbq}. Because $h \in \sC^1([0,1])$, we can expand the ratio $\frac{\mu_Y(y)}{\sigma_Y^2(y)}$ at $y=0$ and at $y=1$ to see 
\begin{align}\label{Taylor1a}
\begin{split}
y \frac{\mu_Y(y)}{\sigma_Y(y)^2}&= \frac{1+\gamma}{2}+ O(y),\quad y\downarrow 0,\\
(1-y) \frac{\mu_Y(y)}{\sigma_Y(y)^2}&= \frac{(\gamma -1) \gamma +\delta }{2 \gamma }+ O(1-y),\quad y\uparrow 1.
\end{split}
\end{align}
{\color{black} Inserting these expansions into $\rho$ in \eqref{rho1} gives}
\begin{align}\label{rhoest1}
\begin{split}
\rho(y) &= e^{\int_y^a \frac{O(x)}xdx} e^{\int_y^a \frac{1+\gamma}{x}dx}\\
&= e^{\int_y^a \frac{O(x)}{x}dx} \Big(\frac{a}{y}\Big)^{1+\gamma},\quad y \in (0,a],
\end{split}
\end{align}
and 
\begin{align}\label{rhoest2}
\begin{split}
\rho(y) &= e^{\int_a^y \frac{O(1-x)}{1-x}dx} e^{\int_a^y \frac{1-\gamma -\frac{\delta}{\gamma}}{1-x}dx}\\
&= e^{\int_a^y\frac{O(1-x)}{1-x}dx} \Big(\frac{1-a}{1-y}\Big)^{1-\gamma -\frac{\delta}{\gamma}},\quad y \in [a,1).
\end{split}
\end{align}
The scale function is defined as
\begin{align}\label{scalefct}
s(y) := \int_a^y \rho(x) dx,\quad y\in (0,1).
\end{align}
{\color{black}From \eqref{rhoest1}, we see $\gamma>0$ ensures that $s(0^+)  = -\infty$ for $\delta \in (-\gamma,0)$. We see from \eqref{rhoest2} that 
\begin{align}\label{s_endpoints}
\begin{split}
&\delta \in (-\gamma,-\gamma^2] \Rightarrow s(1^-)= \infty,\\
&\delta \in (-\gamma^2,0)\Rightarrow s(1^-)< \infty.
\end{split}
\end{align}

{\bf Step 2/3:} Consider the left-end point $y=0$. Because 
$$
y \frac{\mu_Y(y)}{\sigma_Y(y)^2}= \frac{1+\gamma}{2}+ b_1(y)y,\quad b_1(y) :=\frac{2 \gamma ^2 h(y)-\gamma(1+\gamma) +\delta  y h(y)^2}{2 \gamma  (1-y)},
$$
with $b_1(y)$ bounded for $y$ close to zero, the proof of Lemma 3.2 in Guasoni, Larsen, and Leoni (2025) applies also in our case with $\delta \in (-\gamma,0)$ and gives that $y=0$ inaccessible and entrance.  To see that $y=0$ is not attracting,  we use $s(0^+)=-\infty$. Assume to the contrary that $\lim_{t\to\infty}Y_t = 0 $ with some positive probability. Then, the continuous local martingale $s(Y_t)$ converges pointwise to $-\infty$ with positive probability. This is impossible, see, e.g., Proposition V.1.8 in Revuz and Yor (2013).

{\bf Step 3/3:} Consider the right-end point $y=1$. Because $\delta$ appears in the second expansion in \eqref{Taylor1a}, the analysis differs slightly from that given in Lemma 3.3 in Guasoni, Larsen, and Leoni (2025). 
For $\delta \in (-\gamma,-\gamma^2]$, \eqref{s_endpoints} gives $s(1^-) =\infty$ and so Theorem 8A in Helland (1996) gives that the boundary point $y=1$ is inaccessible. For $\delta \in (-\gamma^2,0)$, the representation of $\rho$ in \eqref{rhoest2} ensures  there exists an irrelevant constant $c$ such that 
\begin{align*}
s(1^-) - s(y) &= s'\big(\theta(y)\big)(1-y) \\
&= \rho\big(\theta(y)\big) (1-y) \\
&\ge  c\Big(\frac{1-a}{1-\theta(y)}\Big)^{1-\gamma-\frac \delta \gamma} (1-y)\\
&\ge  c(1-a)^{1-\gamma-\frac \delta \gamma} (1-y)^{\gamma+\frac \delta \gamma},
\end{align*}
where $\theta(y) \in (y,1)$ is produced by the Mean-Value Theorem. Because $h\ge \gamma$, the function $\sigma_Y$ from \eqref{mgmethod1bbq} satisfies
$$
\sigma_Y(y)^2 = \frac{\sigma_D^2 (1-y)^2}{h(y)^2} \le \frac{\sigma_D^2(1-y)^2 }{\gamma^2},\quad y\in [0,1].
$$
Combining these two inequalities gives for $y\in (a,1)$ the following lower bound
\begin{align}\label{scalefctV}
\begin{split}
\frac{s(1^-) - s(y)}{\rho(y)\sigma_Y(y)^2 } &\ge c_2 \frac{(1-y)^{\gamma+\frac \delta \gamma}}{(1-y)^{\gamma +\frac{\delta}{\gamma}-1}(1-y)^2 }\\
 &= \frac{c_2}{1-y },
 \end{split}
\end{align}
for some irrelevant constant $c_2>0$. This lower bound is not integrable for $y$ near 1. Therefore,  for $\delta \in (-\gamma,0)$, Theorem 8A in Helland (1996) gives that the boundary point $y=1$ is inaccessible.

The representation of $\rho$ in \eqref{rhoest2} gives  irrelevant constants
$c_3,C_3>0$ such that
\[
    c_3(1-x)^{\gamma+\delta/\gamma-1}
    \leq \rho(x)
    \leq
    C_3(1-x)^{\gamma+\delta/\gamma-1},
    \qquad x\in[a,1).
\]
In turn, this gives the lower bound 
\begin{align*}
\frac{s(y)}{\rho(y)\sigma_Y(y)^2}  &\ge c_4 \frac{\int_a^y  (1-x)^{\gamma+\frac \delta \gamma-1}dx}{(1-y)^{1+\gamma+\frac \delta \gamma}}= 
 c_4\begin{cases}
 \frac{\gamma  \left((1-a)^{\frac{\delta }{\gamma }+\gamma }-(1-y)^{\frac{\delta }{\gamma }+\gamma }\right)}{(\gamma ^2+\delta)(1-y)^{1+\gamma+\frac \delta \gamma}},\quad \delta \neq -\gamma^2,\\
  \frac{\log \left(\frac{1-a}{1-y}\right)}{(1-y)^{1+\gamma+\frac \delta \gamma}},\quad \delta = -\gamma^2.
 \end{cases}
\end{align*}
This lower bound is not integrable for $y$ near 1. Therefore,  for $\delta \in (-\gamma,0)$, Theorem 8C in Helland (1996) ensures that $y=1$ is natural (i.e., $y=1$ is not entrance).

When $\delta \in ( -\gamma^2,0)$, we see from \eqref{s_endpoints} that $s(1^-)< \infty$.  Proposition 5.5.22(c) in Karatzas and Shreve (1998) and $s(0^+) =-\infty$ give us that $Y_\infty=1$ almost surely. Therefore, when $\delta \in ( -\gamma^2,0)$, $y=1$ is attracting. When $\delta \in (-\gamma, -\gamma^2]$, \eqref{s_endpoints} gives $s(1^-) = \infty$ and so Theorem 8D in Helland (1996) ensures that $y=1$ is not attracting.}

 {\color{black} The functions $\mu_Y$ and
$\sigma_Y$ in \eqref{mgmethod1bbq} are locally Lipschitz. Hence, pathwise uniqueness and a
unique strong solution hold up to the first exit time from $(0,1)$.
The preceding boundary classification shows that this exit time is
almost surely infinite.}

$\endproof$\ \\

\noindent \emph{Proof of Theorem \ref{Main}:} Given Lemma \ref{MainLemma}, the proof of Theorem \ref{Main} is identical to the proof of Theorem 3.6(2) in Guasoni, Larsen, and Leoni (2025) {\color{black}when the traders' optimal state-price densities are adjusted to 
\[
   \hat Z_{1,t}=e^{-\beta_1t}(D_tY_t)^{-\gamma},
    \qquad
    \hat Z_{2,t}=e^{-\beta_2t}\big(D_t(1-Y_t)\big)^{-\gamma}.
\]}
$\endproof$

{\color{black}
\noindent \emph{Proof of Theorem \ref{thm:Prieto}:} 1. Prieto (2013) proves the result for $\beta_1=\beta_2$. The same derivation applies for different time-preference parameters, with the resulting
drift becoming $\mu_Y$ as displayed.

2. We adjust the proof of Lemma \ref{MainLemma}(2) based on 
\begin{align}\label{TaylorPrieto}
\begin{split}
y \frac{\mu_Y(y)}{\sigma_Y(y)^2}&= \frac{1+\gamma}{2}+ O(y),\quad y\downarrow 0,\\
(1-y) \frac{\mu_Y(y)}{\sigma_Y(y)^2}&= {\color{black}\eta}+ O(1-y),\quad y\uparrow 1.
\end{split}
\end{align}
Based on \eqref{TaylorPrieto}, we see 
$s(0^+)=-\infty$ always whereas 
$s(1^-) < \infty$ for $\eta >-\frac12$ and $s(1^-)=\infty$ for $\eta \le -\frac12$.
The rest of the proof only requires minor modifications to the proof of Lemma \ref{MainLemma}(2) and is therefore omitted.

$\endproof$

}

\end{document}